\documentclass[a4paper]{article}

\usepackage{INTERSPEECH2020}
\usepackage{subfigure}
\usepackage{amsmath, bm}

\title{The HUAWEI Speaker Diarisation System for the VoxCeleb Speaker Diarisation Challenge}
\name{Renyu Wang$^*$ \thanks{$^*$Corresponding author}, Ruilin Tong, Yu Ting Yeung, Xiao Chen}
\address{
  Huawei Noah's Ark Lab, China}
\email{ \{wangrenyu1,tongruilin,yeung.yu.ting,chen.xiao2\}@huawei.com }

\begin{document}

\maketitle
\begin{abstract}
This paper describes system setup of our submission to speaker  diarisation track (Track 4) of VoxCeleb Speaker Recognition Challenge 2020.
Our diarisation system consists of a well-trained neural network based speech enhancement model as pre-processing front-end of input speech signals. We replace conventional energy-based voice activity detection (VAD) with a neural network based VAD. The neural network based VAD provides more accurate annotation of speech segments containing only background music, noise, and other interference, which is crucial to diarisation performance.
We apply agglomerative hierarchical clustering (AHC) of x-vectors and variational Bayesian hidden Markov model (VB-HMM) based iterative clustering for speaker clustering.
Experimental results demonstrate that our proposed system achieves substantial improvements over the baseline system, yielding diarisation  error  rate  (DER) of  10.45\%,  and  Jacard  error  rate   (JER) of  22.46\% on the evaluation set. 

\end{abstract}

\section{Introduction}

Speaker diarisation is a task of annotating a conversation into homogeneous segments of same speakers, which is often referred to as ``who spoke when'' problem \cite{reynolds2005approaches, tranter2006overview, anguera2012speaker}.
It is an interesting and challenging research topic due to complex scenarios in real-life conversations.
A diarisation system often serves as a pre-processing step for different speech applications, such as speaker tracking, speech transcription, and meeting summary.

A conventional speaker diarisation system usually consists of two main components, namely speech segmentation and speaker clustering. 
A speech segmentation component usually consists of a voice activity detection (VAD) module and a speaker change point detection module in order to split an audio stream into homogeneous segments. Ideally, each segment should only contain one speaker.
Methods based on generative models and neural discriminant models \cite{chen1998speaker, chen2011learning, wang2017speaker} have achieved considerable performance in various segmentation tasks.
A speaker clustering algorithm is usually belonged to one of the following two categories, bottom-up and top-down methods \cite{anguera2012speaker, evans2012comparative}. 
A classic bottom-up approach such as agglomerative hierarchical clustering (AHC) \cite{han2007robust} tries to merge speech segments iteratively until one or more previously-defined stopping criteria are reached.
In contrast, a top-down approach tries to split speech segments to new speaker clusters according to splitting prerequisites \cite{bozonnet2010lia}. We refer readers to \cite{evans2012comparative} for further comparison of the two approaches.

The recent advances in sequence modeling of deep neural network (DNN) lead to development of supervised speaker diarisation systems with end-to-end DNN models \cite{zhang2019fully, fujita2019end, shafey2019joint}. These end-to-end DNN models perform online speaker diarisation without segmenting and clustering an input audio stream. However, end-to-end methods require large amount of human-annotated or simulated conversations as training data. Preparation of training data which is suitable for real-life scenarios is a main challenge for end-to-end diarisation models. 

System performance of speaker diarisation continues to improve for both conventional systems and end-to-end DNN models. However, lack of a common task has resulted fragmentation of individual research groups focusing on different datasets or domains.
Researches and challenges focusing on diarisation of real-life scenarios \cite{sell2018diarization, ryant2019second} catalyse the development of speaker diarisation systems for practical applications. 

The diarisation track of VoxCeleb Speaker Recognition Challenge 2020 is launched as a satellite event of Interspeech 2020 to facilitate study on speaker diarisation on videos collected ``in the wild'' \cite{chung2020spot}. The dataset includes multiple real-life scenarios such as talk shows, news broadcast, interviews, and vlogs.
The dataset covers large numbers of speakers. Short rapid exchanges cross-talk, applause, background music, noise, and reverberation are common in the recordings, which would degrade diarisation performance. For example, noisy non-speech segments and speech segments contaminated by noise are easily mishandled as new speaker clusters. When these errors accumulate, the quality of final diarisation output could be significantly degraded, and probably unusable. 
Therefore, speech front-end pre-processing is essential for speaker diarisation. Improved speech quality generally leads to higher performance upper-bound for a speaker diarisation system \cite{sun2018speaker}.

In this paper, we describe our system and experimental results of this challenging diarisation task. Our system is still based on the methodology of a conventional speaker diarisation system.
First, we build a baseline system based on energy-based VAD, x-vector speaker representation and an AHC based clustering \cite{snyder2018x}. 
For the advanced system, we add a DNN-based speech enhancement model with long short-term memory (LSTM) architecture as speech pre-processing front-end, and replace the energy-based VAD with a DNN-based VAD. After these modifications, diarisation errors due to mis-classifying noise and music segments as speech segments are reduced significantly, leading to substantial improvement in diarisation performance.

This paper is organized as follows. We present the structure of our diarisation system in Section 2. Experimental results are presented in Section 3, followed by conclusions in Section 4.

\section{System Structure}

\begin{figure}[t]
  \centering
  \includegraphics[width=\linewidth]{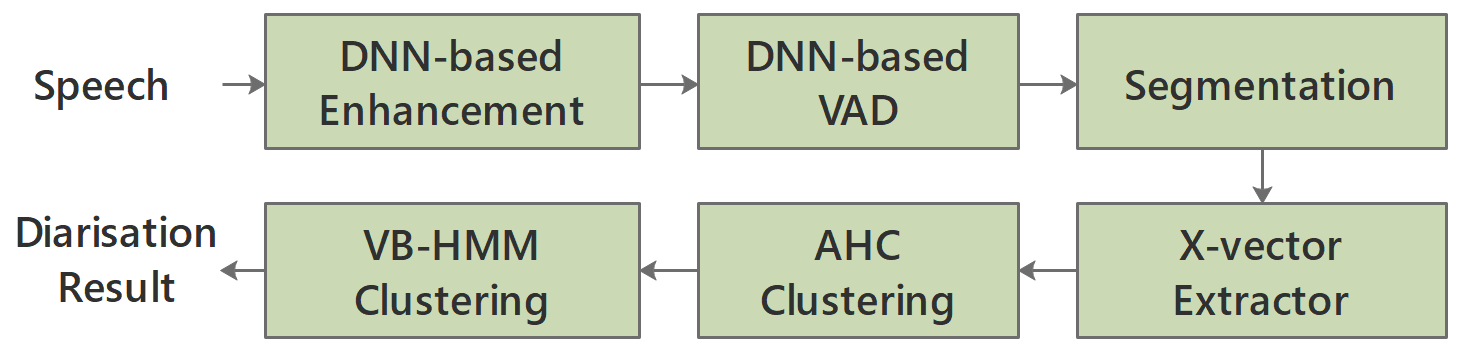}
  \caption{Framework of the Huawei speaker diarisation system}
  \label{fig:speech_production}
\end{figure}

Our speaker diarisation system consists of several components including speech enhancement, voice activity detection, speaker feature representation, speaker segmentation and clustering. 
The overall structure of the proposed speaker diarisation system is shown in Figure \ref{fig:speech_production}. 

\subsection{Neural network based speech enhancement}

An effective speech pre-processing front-end should be able to remove most of background noise and music, and retains most of speaker information.
Due to limitation of model assumption, traditional speech enhancement methods such as Winener filtering \cite{anguera2012speaker}, and minimum-mean  square  error (MMSE) estimators \cite{ephraim1984speech} perform poorly on non-stationary noise, and induce artifact to speech spectrum. These lead to loss of speaker information and degrade performance of a speaker diarisation task. Therefore, we adopt a DNN-based speech enhancement model with advanced LSTM architecture with specially designed hidden layers and multiple learning targets of predicting both log-power spectra (LPS) feature and ideal ratio mask (IRM).
In order to retain most of speaker information, we apply densely connected progressive learning strategy proposed in \cite{gao2016snr}. The input feature and the estimated target are spliced together for learning next targets of higher signal noise ratio (SNR). We apply minimum mean square error (MMSE) criterion for parameter optimization.

\begin{figure}[tt]
  \centering
  \subfigure[Original speech segment]{
  \includegraphics[width=\linewidth]{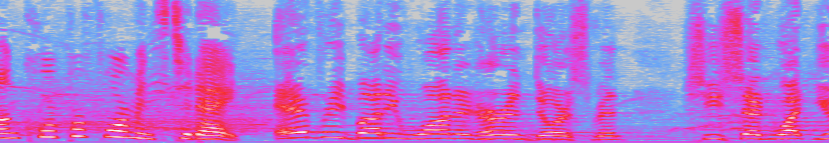}
  }
  \quad
  \subfigure[Enhanced speech segment]{
  \includegraphics[width=\linewidth]{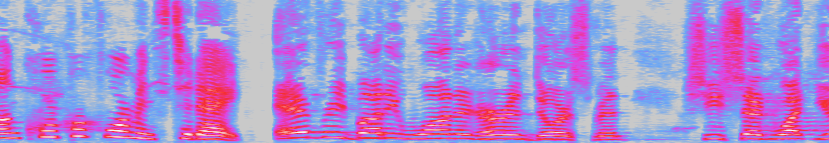}
  }
  \caption{Comparison of spectrograms with background music and applause for the proposed DNN-based speech enhancement}
  \label{fig:speech_enhancement}  
\end{figure}

A sample speech segment selected from the evaluation set are shown in Figure \ref{fig:speech_enhancement}. We compare spectrograms before and after speech enhancement.
In the original spectrogram, speaker's speech was accompanied with cheers, applause and music. After speech enhancement, a significant amount of interference has been removed, but spectral components of speech signals are generally retained.

\subsection{Neural network based VAD}

The recordings from this challenge is collected ``in the wild''. There are various types of noise segments of different lengths between different speakers. These noise segments are easily mishandled as new speaker clusters if they are not correctly removed. A well-trained VAD is essential for this requirement.
The core part of the VAD module is a deep neural network (DNN) for binary classification of speech and non-speech frames. The DNN model consists of two 2D-convolution layers, and three time-delay layers. Left and right context of the time-delay layers are asymmetric, in order to reduce latency during streaming inference. We also apply AM-softmax \cite{wang2018additive} at DNN output for better frame-wise discriminative training. 
We apply 128-dimensional Mel-frequency filterbanks as input feature. The input feature is normalized globally. We perform force alignment on training data with our automatic speech recognition (ASR) system to obtain frame-wise speech and non-speech labels. 

We further apply post-processing on the neural network based VAD model output. We apply speech smoothing to remove speech chunks shorter than 200~ms, and apply speech segmentation when silence is longer than 500~ms.

\subsection{Speaker feature representation}

After segmentation of recordings from VAD module, we get speech segments of varying lengths. A good speaker feature representation in a diarisation system should be able to capture speaker discriminative features in a short speech segment and more accurate speaker feature representation in a long speech segment. X-vector as our speaker feature representation can fulfill this requirement \cite{snyder2018x}. X-vector is extracted from a well-trained time-delay neural network (TDNN) model according to the VoxCeleb recipe of the Kaldi toolkit \cite{povey2011kaldi}. We apply 40-dimensional Mel-frequency filterbanks as speaker input feature. The TDNN model consists of three components. First, there is a feature learning component, which consists of  5 time-delay layers for higher order representation of speaker filterbank feature. The slicing parameters of the 5 time-delay layers are
$ \{ t-2, t-1, t, t+1, t+2 \}, \{ t-2, t, t+2 \}, \{ t-3, t, t+3 \}, \{ t \}, \{ t \} $ respectively.
Then, there is a statistical pooling component, which computes mean and standard deviation of the higher order speaker representation of a given speech segment.
The final component is a speaker classification module which consists of two fully-connected layers and a softmax output layer. The size of the softmax output corresponds to number of speakers in training set. The x-vector are extracted from the penultimate fully-connected layer once the network is well-trained.

Considering complex noise and reverberation conditions in the Challenge, we apply data augmentation based on the pipeline of Kaldi SRE16 recipe \cite{snyder2018x} during training. We augment the training data by mixing with various music and noise signals and convolving with the different room impulse responses for reverberation \cite{ko2017study}.

\subsection{AHC and VB-HMM clustering}

We apply agglomerative hierarchical clustering (AHC) and variational Bayesian (VB) based iterative clustering \cite{landini2020but} for speaker clustering in our diarisation system.
The x-vector extracted from speech segments after VAD are first clustered by AHC, with log-likelihood scores from probabilistic linear discriminant analysis (PLDA) as similarity metric. PLDA metric is widely used in speaker verification.

After the initial assignment from AHC, we apply variational Bayesian hidden Markov Model (VB-HMM) at x-vector level to further cluster x-vectors \cite{diez2019analysis}.
The x-vectors are first projected to low-dimensional discriminative space with linear discriminant analysis (LDA).  Variational Bayesian hidden Markov model (VB-HMM) inference is performed iteratively for refining assignment of x-vectors to speaker clusters for more accurate speaker distribution.

\section{Experimental Results}

In this section, we describe datasets for building our diarisation system. We also report the results of our system in terms of diarisation error rate (DER) and Jaccard error rate (JER).

\subsection{Dataset description}

The diarisation task in the VoxCeleb Speaker Recognition Challenge is an open-set track, which allows the use of extra data. Here, we introduce all the datasets used for building each part of our system.

The neural network based speech enhancement system is trained with CHiME-3 dataset \cite{barker2015third}. The speech signals  from the near-field microphone (channel 5) are  mixed with background noise with SNR levels of -5dB, 0dB, 5dB.

For neural network based VAD, we apply LibriSpeech \cite{panayotov2015librispeech} and CommonVoice \cite{ardila2019common} datasets as training data. We augment the training data by adding additive noise with SNR of 0-20dB from AudioSet dataset \cite{gemmeke2017audio}.

For x-vector model training, we choose VoxCeleb corpus \cite{nagrani2017voxceleb, chung2018voxceleb2} containing VoxCeleb 1 and 2 with 1.2 million speech utterances from 7146 speakers.
Data augmentation is performed with MUSAN Corpus \cite{snyder2015musan} and RIRs from the AIR dataset \cite{ko2017study}.

Details of development set and evaluation set of VoxCeleb Speaker Recognition Challenge can be found at \cite{chung2020spot}. Note that we perform speaker diarisation only with audio stream of the dataset.

\subsection{Evaluation metrics}

The diarisation system is evaluated with two metrics, including diarisation error rate (DER) and Jacard error rate (JER). The two metrics compare the system result with reference human annotations.

Diarisation error rate is a percentage measure of total duration of speaker time slots which does not attribute to target speakers correctly. DER considers 3 types of error. FA is total system duration which does not attribute to reference speakers. MISS is total reference speaker duration which does not attribute to system speakers. ERROR is total reference speaker duration attributed to wrong speakers. DER is defined as,

\begin{equation}
DER = \frac{FA+MISS+ERROR}{TOTAL}.
\end{equation}

Jaccard error rate is based on the Jaccard index and is defined as the ratio between intersection and union of system and reference speaker duration. FA is total system speaker duration which does not attribute to reference speakers. MISS is the total reference speaker duration which does not attribute to system speakers. Total JER is an average of JERs of all speakers and is defined as, where N is total number of speakers,
\begin{align}
JER_{spk} &= \frac{FA+MISS}{TOTAL}\\
JER &= \frac{1}{N} \sum_{spk} JER_{spk}.
\end{align}

\subsection{Results}

First, we evaluate our baseline speaker diarization system which is based on energy-based VAD, x-vector based speaker representation and AHC based speaker clustering. For the baseline system, we use VoxCeleb dataset to train a TDNN model as x-vector extractor. The results matches the official performance of the Challenge \cite{chung2020spot}, as shown in Table \ref{tab:example} and Table \ref{tab:example2}. 

Then we study the performance of VB-HMM based clustering. As shown in first two lines of Table \ref{tab:example} and Table \ref{tab:example2}, the VB-HMM iterative clustering method improves the performance of both development set and evaluation set significantly. This gives us a great confidence in solving such a hard ``in the wild'' problem by incorporating VB-HMM method. 

After replacing the energy-based VAD to our neural network based VAD, we get significant performance improvement. When tuning the VAD threshold from 0.8 to 0.5 on the development set, we attain the best results at the threshold of 0.5. We keep this threshold for the evaluation set.

The results confirm that VAD is a critical component for a diarisation system. We successfully remove music, noise and other daily-life non-speech segments with neural network based VAD. These interference could lead to inevitable mistakes in the clustering back-end. An effective VAD helps to achieve higher performance upper bound of a diarisation system. However, we notice that the performance of the development set is sensitive to the VAD thresholds. This may be an interesting topic for further study. 

Applying neural network based speech enhancement (NN-enhanced) further improves the performance slightly. Speech enhancement should be able to minimise the impact of noise and retain more speaker discriminative feature in speech segments. Therefore, we further fine-tune the AHC threshold from 0.0 to 0.8 in speaker clustering stage to allow more aggressive clustering. The performance is further improved. 

We perform diarisation on the evaluation set with the best-performed parameters in the development set. Finally, we get DER of 10.45\%, and JER of 22.46\%. The performance improves substantially over the baseline method. We have submitted the results to the official Challenge submission system. 

\begin{table}[t]
  \caption{Performance on development set}
  \label{tab:example}
  \centering
    \scalebox{1}{
    \begin{tabular}{p{5cm}cc}
    \toprule
    \textbf{methods}      & \textbf{DER}  & \textbf{JER}              \\
    \midrule
    AHC  (baseline)                                   & $23.09$ & $47.61$           \\
    AHC + VB-HMM                              & $17.06$ & $31.07$           \\
    AHC + VB-HMM + NN-VAD (thr-0.8)                     & $16.87$ & $34.85$           \\
    AHC + VB-HMM + NN-VAD (thr-0.7)                     & $12.81$ & $30.54$           \\
    AHC + VB-HMM + NN-VAD (thr-0.6)                     & $11.54$ & $29.30$           \\
    AHC + VB-HMM + NN-VAD (thr-0.5)                     & $10.58$ & $27.77$           \\
    AHC + VB-HMM + NN-VAD (thr-0.5) + NN-enhanced       & $10.06$ & $26.73$ \\
    AHC (thr-0.8) + VB-HMM + NN-VAD (thr-0.5) + NN-enhanced    & $\bm{9.46}$ & $\bm{25.11}$          \\
    \bottomrule
  \end{tabular}
  }
\end{table}

\begin{table}[t]
  \caption{Performance on evaluation set}
  \label{tab:example2}
  \centering
  \scalebox{1}{
    \begin{tabular}{p{5cm}cc}
    \toprule
    \textbf{methods}      & \textbf{DER}  & \textbf{JER}              \\
    \midrule
    AHC  (baseline)                                   & $21.74$ & $51.89$           \\
    AHC + VB-HMM                                        & $18.03$ & $36.34$           \\
    AHC + VB-HMM + NN-VAD (thr-0.5)                              & $11.35$ & $\bm{21.62}$           \\
    AHC (thr-0.8) + VB-HMM + NN-VAD (thr-0.5) + NN-enhanced      & $\bm{10.45}$ & $22.46$           \\
    \bottomrule
  \end{tabular}
  }
\end{table}

\section{Conclusions}

This paper presents the development of our diarisation system for diarisation task of  VoxCeleb Speaker Recognition Challenge 2020.
A neural network based speech enhancement model is applied for speech pre-processing for reducing background interference. We believe that the improved speech enhancement model helps to retain speaker information in the enhanced speech segments.
A well-trained neural network based VAD is applied to identify non-speech segments of music, background noise and other daily-life interference. The  neural network based VAD also helps to obtain accurate speech boundaries for the back-end clustering algorithms. We notice that the main performance improvement of the diarisation system is contributed by the neural network based VAD.
We apply a TDNN based x-vector system for speaker feature extraction followed by AHC for initial speaker clustering. Variational Bayesian hidden Markov model (VB-HMM) clustering at x-vector level further improves the clustering performance by locating more accurate speaker boundaries.
We also apply various data augmentation to increase diversity of training data for each component. By combining these strategies, we achieve substantial improvement on the diariation results in terms of both DER and JER.

Speaker diarisation is a hard task as a system has to solve multiple problems in complex acoustic environments and different applications. In the future, we aim to further improve diarisation performance by investigating overlapped speech detection and automatic threshold selection for real-life scenarios.

\bibliographystyle{IEEEtran}

\bibliography{mybib}

\end{document}